\documentclass[12pt]{article}

\usepackage{amsmath}
\usepackage{graphicx}
\usepackage{enumerate}
\usepackage{natbib}
\usepackage{url} 

\date{}

\usepackage{amsmath} 

\usepackage{graphicx} 

\usepackage{setspace} 

\usepackage{csquotes} 

\usepackage{amssymb} 

\usepackage{xfrac}

\usepackage{tabularx}

 \usepackage{hyperref} 

\usepackage{nicematrix,tikz} 

\usetikzlibrary{fit}

\usepackage{tikz}

\newcommand{\Expect}{{\rm I\kern-.2em E}}

\usepackage{floatpag}

\DeclareMathOperator\arctanh{arctanh}            

 \usepackage{nicefrac}

\usepackage{tcolorbox}

\makeatletter
\newcommand\footnoteref[1]{\protected@xdef\@thefnmark{\ref{#1}}\@footnotemark}
\makeatother

\begin{document}

\title{Spatio-Causal Patterns\\  of Sample Growth}

  \author{Andre F. Ribeiro\footnote{University of Sao Paulo, Sao Carlos, SP, 13560-970,  Brazil.}\\ ribeiro@alum.mit.edu}

\maketitle

\begin{abstract}
Different statistical samples (e.g., from different locations) offer populations and learning systems observations with distinct statistical properties. Samples under (1) 'Unconfounded' growth preserve systems' ability to determine the independent effects of their individual variables on any outcome-of-interest (and lead, therefore, to fair and interpretable black-box predictions). Samples under (2) 'Externally-Valid' growth preserve their ability to make predictions that generalize across out-of-sample variation. The first promotes predictions that generalize over populations, the second over their shared uncontrolled factors. We illustrate these theoretic patterns in the full American census from 1840 to 1940, and samples ranging from the street-level all the way to the national. This reveals sample requirements for generalizability over space and time, and new connections among the Shapley value, counterfactual statistics, and hyperbolic geometry.
\end{abstract}

\def\spacingset#1{\renewcommand{\baselinestretch}%
{#1}\small\normalsize} \spacingset{1}

\vfill


\section{Introduction} 


Large scale and high-dimensional geospatial datasets currently offer rich opportunities for predictive and Geo-AI applications ~\citep{10.1145/3468791.3472263,10.5555/3207692.3207707,10.1145/3516523,Shekhar:2020ww} (e.g., disease incidence, ecological behavior, electoral results, crime occurrence, economic growth, recommendation systems). While it is common practice to train regression and classification models in data collected across distinct locations, little is known about how  their out-of-sample accuracy ('predictiveness') and biasedness (e.g., black-box 'fairness') ~\citep{10.1613/jair.1.12228,Tiddi:2022vw} are expected to change across spatial extensions. The first indicates whether predictions derived from the sample will be close to their true values for a population in conditions different from at the time of data collection (i.e., whether they will 'generalize'), and the latter whether they will systematically favor individual populations (e.g., as result of their smaller sizes, unobserved variables, or other failures in sample selection). Understanding these issues is important because they allow us to answer crucial questions for collected samples: Can predictions made for a given population with data from one location be used in others?  Does collecting larger samples, or data from distinct locations, improve prediction accuracy for that first population? We first formulate theoretic functions describing fairness-generalizability tradeoffs across space, revealing their connections to hyperbolic geometry and theoretic experimental designs. We then consider 100 years of the American census (and all variables in the census) as case study. For each cross-section (decade), we consider the important task of predicting economic growth for over 60K individual locations under increasing spatial samples. We demonstrate how (1) generalizability tradeoffs evolve across spatial levels, and (2) repeat the validation of generalizability limits derived in ~\citep{ribeiro-ev} for the spatial domain, and with the current census micro-data. 

Let $S : X_m \rightarrow [0,1]$ describe any learning system or agent using an input sample $X_m$ with $m$ variables to derive a classification decision, $S(X)$. Our central goal is to formulate how the generalizability of these systems changes across space (i.e., to what extent  a model assembled in a location will hold for others), and, thus, to identify a parametric functional form $\mathcal{F}$ that can describe accuracy bounds across possible $S$, 

\begin{align} \label{eq-problem}
\max \Big\{ \textrm{ACC}\Big( S; X \Big[\, x_0, \; d_{x_0}\,\Big]\Big) \Big\}= \mathcal{F}\Big(\, d_{x_0}\, \Big),
\end{align} 

where $\textrm{ACC}$ indicates the accuracy of models trained in samples $X[x_0;\, d_{x_0}]$ encompassing all  observations at distances less than or equal to $d_{x_0}$ from $x_0$.  The \textbf{specific way in which $\mathcal{F}$ changes across space offer limits and opportunities to algorithmic and agent learning systems and their performance}. Strict bounds on the uncertainty of predictions afforded to algorithms or agents using local data can have a profound impact on the usefulness, scope, and quality of their strategic decisions or the recommendations they offer. A non-decreasing $\mathcal{F}$ indicates that $S$ is able to generate models that are accurate across locations, while a decreasing function indicate that learning fails to generalize across locations. A complementary, but related, issue would be to what extent $S$ would be able to identify the effect (or importance) of individual variables to prediction based on the same sample. 


To illustrate these two natural sample characteristics, consider a set of binary attributes $X_4=\{a,b,c,d\}$ observed across US locations, such as, respectively, recorded presence of crime, police stations, banks and ice-cream shops. The behavior of these entities are possibly interconnected, which implies that any calculated statistic $y  \in \mathbb{R}$ over $a$ (e.g., crime incidence) is in fact an statistic over $y( a \; | \; b, c, d)$. Because of this, we say that this local observation has low External Validity (EV), since any changes in factors $\{b, c, d\}$ (or the many other factors that can conceivably affect crime), can invalidate the statistic. At the same time, because banks often appear together with ice-cream shops in commercial and affluent neighborhoods, we are not sure whether their presence plays any essential role when predicting crime incidence (i.e., whether they have only a spurious relationship to crime). Because of this, we say that observations are confounded (CF) in this sample. How can these two issues be addressed and quantified? Comparison of crime incidence between this location and a second with banks but no ice creams shops, everything else constant, would lend evidence to the fact that ice cream shops are not driving crime up. These types of ideal what-if statements, where the effect of an outcome is observed under a single or small difference (while holding other factors constant), are called counterfactual statements ~\citep{Morgan:2007aa,Rubin:2005aa}.  In the hypothetical case where all conditions like these can be observed, the problem of determining whether a factor is relevant to prediction is fairly easy to solve. Many of these set of sample combinatorial conditions have been formulated mathematically, for example, in the study of experimental designs ~\citep{Montgomery:2001aa}. The more challenging, and practical, aspect of this problem is, however, the case of unobserved conditions: often the relevant factors that change across locations, and samples, are both not observed and held constant. The issue of unobserved confounding is particularly serious, as the statistics $y$ calculated from the sample might then also be influenced by 'exogeneous' or unobserved variation that cannot be easily controlled or discounted by typical regression and effect estimation methods ~\citep{Scholkopf:2021aa,Morgan:2007aa}. Although in these conditions effect observations are only flawed or noisy counterfactual observations, we still refer to them as 'counterfactual' observations for short. We study these problems by studying spatial sample growth: we start with a sample with a single unit (all conditions unobserved), we then progressively add units to the sample at increasingly larger distances to the first, progressively decreasing its number of unobserved conditions. We consider their relation to $\textrm{ACC}\Big( S; X \Big[\, x_0, \; d_{x_0}\,\Big]\Big)$, and, in particular, how EV and CF change, as result, across scales. 

The two previous problems reflect two key, but distinct, learning problems ~\citep{Kleinberg:2015aa}: supervised out-of-sample prediction, and factor effect (or importance) estimation. Supervised prediction focuses on making accurate predictions of an outcome in unseen data using an input training sample, while factor effect estimation aims to measure the relevance of specific factors for prediction and model selection. Solutions to the former have become associated more closely with the 'Machine Learning' moniker, and the latter with traditional causal effect estimation. The first problem stresses the construction of predictive models, and the the second interpretable and unbiased models. More recently, they have been combined, and the latter has also been studied when using, exclusively, the output of black-box supervised predictors ~\citep{10.1613/jair.1.12228}. These problems are, however, closely connected ~\citep{Kleinberg:2015aa}, as complete and correctly specified models lead to accurate predictions. We will demonstrate that combinatorial properties of samples impact these two traditional problems differently and reveal tradeoffs across samples and locations.  We first preview the two central contributions of the proposed framework. After this summary, we relate the approach to known black-box importance estimation and causal effect estimators used in practice and formulate the proposed combinatorial and geometric relations in further detail.  

\begin{figure} 
\centering 
\includegraphics[width=1\linewidth]{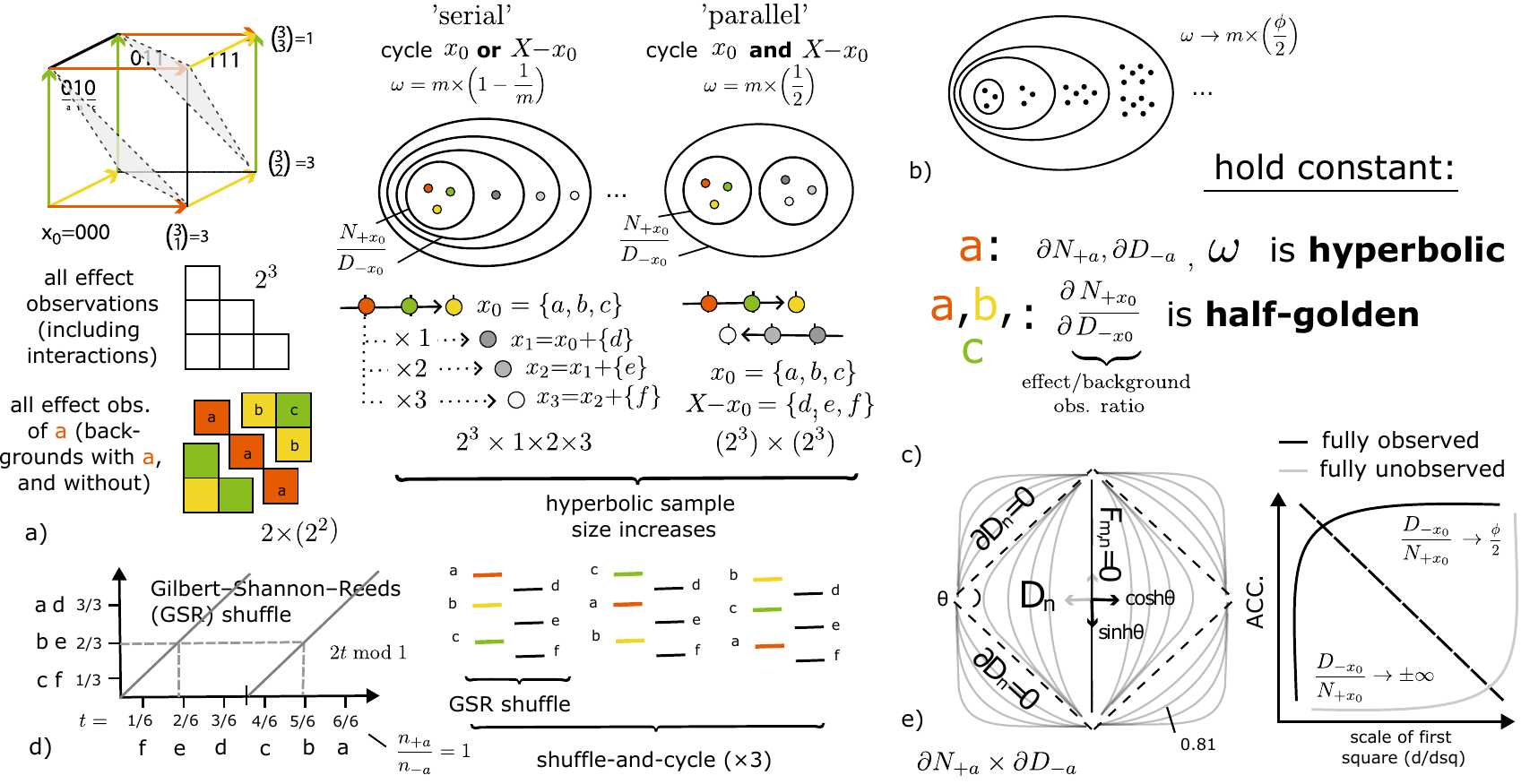}\\ 
 \caption{\footnotesize \textbf{(a)} Can counterfactual effect observations made in one location be used in another (are they externally valid, EV)? Can their independent effects or relevance to prediction be distinguished from others (are they unconfounded, CF)? a $m$-dimensional hypercube and Pascal triangle with $m$ rows portrays the full set of counterfactual effect observations with $m$ factors in a sample $X_m$ ($m=3$),  a $m{\times}m$ Latin-Square ('square') portrays all effect observation backgrounds, more counterfactual effect observations increase guarantees over the generalizability and bias of samples' effect observations,  \textbf{(b)} 'serial' and 'parallel' interleaving of in and out of sample factors during sample growth and their expected sample sizes and growth rates, \textbf{(c)} sample sizes in a sweep follow hyperbolic forms, with a half-golden rate of increase for the high-dimensional case, \textbf{(d)} illustration of Gilbert–Shannon–Reeds (GSR) shuffling as a mechanism to increase generalizability (EV) in partially-observed samples, and a shuffle-and-cycle strategy as an alternative which also guarantee decreases in factor effect confoundness (CF), \textbf{(d)} hyperbola for samples with distinct $\nicefrac{N_{-a}}{N_{+a}}$ (left) and expected accuracy for samples with small or large numbers of unobserved factors (right). }\label{fig-intro} 
\end{figure} 



\subsection{Sample Growth Processes} 

Let $X=\{a,b,c,..., [m]\}$ be a set of (observed or unobserved) binary factors\footnote{where $[m]$ is the $m$-th factor in sample $X$.} characterizing a population $x$, $x \in [0,1]^m$,  and $y(x)$ be a measurement over the population, $y(x) \in \mathbb{R}$. Consider a sample 'growth' process where we start with a fixed sample unit $x_0$ and, as we observe each new unit, we also observe their differences from $x_0$, in respect to both $x$ and $y$. For example, by observing a unit $x_1$ with $x_0-x_1 = \{a\}$, we are also making a single observation for the effect of $a$ in $y$. The growth 'space' for this collected sample is then the imaginary space that contains all of the conceivable ways in which we could have assembled $X$ from any one of its individual sample units $x_0$ (\textit{Sect. Combinations, Permutations and Partial Permutations}). It is often unobservable. This is a problem in samples where their factors cannot be assumed to be (1) independent, or (2) in-sample. That's because (effect) observations can then be contingent on (1) the order or history of the growth process, or (2) out-of-sample factors ~\citep{ribeiro-ev}. In practice, \textbf{two timescales determine the statistical properties of the sample growth process}: the rate $\partial N_{+a}$  at which individual factor differences are observed (i.e., when their counterfactual effects can be observed), and the rate $\partial N_{-a}$ at which the 'backgrounds' in which they are observed change. If the relationship between these two timescales is such that effects are observed under a large number of backgrounds, then we can be more confident about their generalisability (i.e., that effect observations will likely be reproducible in future backgrounds). At the same time, if observations are such that the same populations are observed under the same backgrounds, we can be more confident about their unconfoundness (i.e., that effects observed reflect the same amount of variation across all sample populations). 

 More specifically, a background $D_{-a}$ of an effect observation of factor $a$ from a set $\mathcal{D}(X{-}\{a\})$, $D_{-a} \in \mathcal{D}(X{-}\{a\})$, is the instantaneous condition in which the effect is observed, $\Delta y[\, a \, |\, D_{-a}\,]$. We will consider definitions where  $\mathcal{D}_{-a}$ corresponds to the set of all possible values over the set of other factors, $\mathcal{D}(X{-}\{a\}) =  X{-}\{a\}$, or their permutations\footnote{where $\Pi(X)$ is the set of all permutations of the set $X$.}, $\mathcal{D}(X{-}\{a\}) = \Pi(X{-}\{a\})$, and when $X$ contains both observed and unobserved factors.  For illustration, imagine all ways we can observe backgrounds across growth trajectories of the previous example, $X=\{a,b,c,d\}$. In order to consider effect observations for $a$, we must 'insert' this individual factor in all its possible $3!=6$ backgrounds observable during growth. After each insertion, we can then observe changes $\Delta \hat{y}(a)$ in the outcome-of-interest, $y$, to understand $a$'s effect on $y$.  Another way of saying this is that we need to keep $a$ constant while cyclically permuting all other sample factors, Fig.\ref{fig-intro}(a). That is, we are called to observe the effect of $a$ under the cyclic permutations across each $d\leq m-1$ sample growth step,
 
 \begin{align}
 \tau(x_d) : \{a,b,c,... [d]\} \rightarrow \{b,c,... [d], a\},
 \end{align}
 
whose iteration  $\tau(x_d), \tau^2(x_d), ... \tau^d(x_d)$ have a shifting action in the original permutation and generates the cyclic group\footnote{The cyclic group generated by $\tau$ is related in an obvious way to the set of all permutations of $m$ elements (i.e. to the $S_m$ symmetric group), where $\tau^i(x)$, $i < d$, correspond to one partial permutation, $\tau^i(x) \in S_m$, for all\footnotemark $\;x \in \mathcal{P}(X)$ (\textit{Sect. Combinations, Permutations and Partial Permutations}).}\footnotetext{\label{foot-power}where $\mathcal{P}(X)$ is the power-set of $X$.} of order $d$ ~\citep{Diaconis:1983uo}.  Whether we require $a$ to be inserted in all background values for factors other than $a$ (a single cyclic background permutation $\tau(x)$ across $d$), or all values and their ordering (all cyclic permutations $\tau(x)$ across $d$), will change the guarantees we can make in respect to the EV and CF of effect observations. At the limit, \textbf{these two cases correspond to $1$ or $m-1$ iterations of the recursive definition of a factorial}, $m!=m\times (m-1)!$ 


  

 
 


More generally, we think of $X{-}x_0$ as the unobserved factors of the subsample $x_0 \subseteq X$ of the (possibly unknown) complete sample $X_m$, and we use high caps to indicate counts over unique values. Define, therefore, variables $N_{+x_0} = |\mathcal{P}(x_0)|$ indicating the number of unique in-sample effect observations\footnoteref{foot-power} in $x_0$, and $D_{-x_0}(X) = |\mathcal{D}(X{-}x_0)|$ their out-of-sample backgrounds.  Fig.\ref{fig-intro}(a) illustrates two standard ways of visualizing the former set of $2^{|x_0|}$ in-sample effects: as the number of edges of an hypercube of dimension $|x_0|$ or sums of Pascal triangle's $|x_0|$-th row. The total number of counterfactual effect observations (i.e., all effects under different backgrounds) for the factors $x_0$ is then given by

\begin{equation}\label{eq-vol}
0 < N_{+x_0} \times D_{-x_0}(X) \leq 2^{|x_0|} \times (m{-}|x_0|)!,
\end{equation}

which is large, $D{-x_0}(X)\gg N{+x_0}(X)$, in the factorial-based definition of a 'background'. A key question then becomes, as sample dimensions grow, what is the \textbf{asymptotic number of backgrounds that effects will typically be observed under}?


\subsection{Combinatorial Shuffling (Accuracy)} 

To illustrate how models trained in the samples with the previous characteristics can have their accuracy constrained, we can relate the previous growth process to a traditional combinatorial randomization scheme such as  Gilbert–Shannon–Reeds (GSR) shuffling ~\citep{diaconis1988group,Diaconis:1983uo}.  Like before, start with a sample containing only one population $x_0 \subseteq X_m$ (defined by a set of binary attributes). Let a population with the same attributes as $x_0$ be represented by the string containing only $1$ values (for factors $a,b,c...$). A string containing differences can then be written as $\sigma(x_0) = \left(\sigma_{1},\sigma_{2},\sigma_{3},\ldots \right)$, where each string position take binary values $\sigma_t \in \{0,1\}$. We are interested in the cyclic shifting operator in this representation.  The action of the operator on a string is $\tau(\sigma_{t},\sigma_{t+1},\sigma_{t+2},\ldots) = (\sigma_{t-1},\sigma_{t},\sigma_{t+1},\ldots)$, that is, each string's position is shifted over by one to the left. Considering effect observations, each such operation reveals the effect of a new single variable, and correspond to a counterfactual observation of effect. We can apply the same representation to out-of-sample, or yet unobserved, factors, leading to a second sequence $\sigma_{-1},\sigma_{-2},\sigma_{-3},\ldots$ until we observe all factors. Sample growth can then be represented as the bi-directional string with positive values in-sample and negative out-of-sample, $\sigma = \left(\ldots,\sigma_{-2},\sigma_{-1},\sigma_{0},\sigma_{1},\sigma_{2},\ldots \right)$.  In an increasing spatial sample, the current scale correspond to the zero-index string position. Similar to the GSR, this bi-directional string may be represented, in turn, by two real numbers $0\le x,y\le 1$ as

\begin{align} 
x(\sigma)=\sum_{t=0}^\infty \sigma_{-t} 2^{-(t+1)}, \quad y(\sigma)=\sum_{t=0}^m \sigma_{t+1} 2^{-(t+1)},
\end{align}



The shifting action $\tau$ for two separate strings is known as a dyadic transformation, which can be thought as a 'folding' or 'shuffling' operation between them - mapping each distinct $x$ to a single and distinct $y$ in each iteration. In this case, the transformation is between one set of factors, $x_0$, and one possible background, $\partial \mathcal{D}(X{-}x_0)$. The right diagram in Fig.\ref{fig-intro}(d) illustrates the result of a GSR shuffle for a 6-letter example with half variables observed. With each shuffle, each in-sample factor effect, of $\{a,b,c\}$ (colored bars), is observed under different backgrounds (e.g., the effect of $a$ is observed under the background of $d$ and of $b$ under $e$). We would therefore expect the generalizability (EV) of effect observations to increase with each such operation.  There are two practical problems with this scheme, however. First, since each factor effect is observed in a different background, each effect observation is confounded by a different factor (e.g., the observed effect of $a$ reflects the influence of unobserved factor $d$). Second, GSR shuffling only shuffles factors under a 1-step markovian assumption. In complex biological and economic systems, for example, higher-order interactions are the rule and not the exception ~\citep{Breen2012,Battiston:2021wu}. Truly shuffling observations then often require systematic permutation of backgrounds of size larger than $1$. Fig.\ref{fig-intro}(d) also illustrates an alternative 'shuffle-and-cycle' scheme, where each shuffle step is followed by $m$ cyclic permutations $\tau, \tau^2, ..., \tau^m$ of in-sample factors. This alternative has three advantages: (1) each factor is observed under every background, and thus effects are confounded in equal proportions across sample populations (thus allowing us, for example, to 'factor-out' these effects more easily ~\citep{ribeiro-ev}), (2) every effect is now observed for the same permutation of unobserved factors ($\{d,c,e\}$ in the figure), and thus generalizability increases at a common rate for all factors and populations, and without markovian assumptions, (3) this is a \textbf{limiting process for every sample} (\textit{Sect. Combinations, Permutations and Partial Permutations}). Each such shuffle-cycle operation will be visualized in this article with a Latin-Square (whose rows show a cyclic-permutation of a starting permutation of observed factors), Fig.\ref{fig-intro},\ref{fig-intro2}(a).

\subsection{Hyperbolic Geometry (Sample Sizes)} 

 To better understand the relationship between sample sizes and out-of-sample performance, we need to consider the sample sizes required to generate all possible ways of interleaving samples' in and out (ordered) factors.  Fig.\ref{fig-intro}(a) outlines two equivalent sample growth patterns that achieve this ('serial', 'parallel'). In the first example, each new background is interleaved with all previous effects 'serially'. For a first out-of-sample factor, $d \in X{-}x_0$, this requires observation of the $N_{+x_0}{=}3$ effects, $x_0{=}\{a,b,c\}$, under the new condition, then $N_{+x_0}{+}1$ effect observations, $x_1{=}\{a,b,c,d\}$, etc. In the second example, several out-of-sample factors are interleaved with several in-sample simultaneously. Effects for $x_0$ are observed, at first, under backgrounds $\{d,e,f\}$, then $\{f,d,e\}$, etc.   Both strategies lead to a geometric series of background enumeration for individual effects analogous to the factorial, but at different rates. Sample sizes in both cases can be described by the hyperbola,
 




\begin{equation*}
\setlength{\abovedisplayskip}{0pt} 
\setlength{\belowdisplayskip}{0pt} 
        \noindent 
\renewcommand{\arraystretch}{0.1}
          \begin{tabularx}{\linewidth}{>{\hsize=0.9\hsize\linewidth=\hsize}X 
>{\hsize=0.1\hsize\linewidth=\hsize}X} 
\begin{equation} \label{eq-intro-hyperbolic}
\Big(\frac{\omega}{\partial N_{+x_0}^2}\Big) - \Big( \frac{\omega}{\partial D_{-x_0}^2}\Big) = 1.
\end{equation} &  
  \begin{equation*} 
(\omega \textit{ const.}) 
\end{equation*}
\end{tabularx} 
    \end{equation*} 

The equation expresses that, for each new in-sample factor, $\partial N_{+x_0}$, we are able to re-measure their effects in each new unobserved background, $\partial D_{-x_0}$, thus increasing their generalizability (EV).  The quantity $\omega$ describes, in turn,  the 'speed' in which EV is expected to increase for individual in-sample populations. Sample sizes follow known exponential, $(1-\nicefrac{1}{m})^t \to e^{-1}$, and binomial, $2^{-1}$, growth rates in these cases, $0 < t \leq m$ ($m \gg 1 $), Fig.\ref{fig-intro}(a).

When all variables are observed across the same number of backgrounds (i.e., in 'balanced' samples\footnote{this is analogous to notions of balancedness in experimental designs ~\citep{Montgomery:2001aa} but require milder conditions than equal-size populations, being observable in multi-frequency and multi-scale processes, and being observed in real-world systems, as demonstrated in \textit{Sect. Results}.}) across all its $m$ factors, samples following Eq.(\ref{eq-intro-hyperbolic}) have sizes, $n_{x_0}$, increasing according to a Fibonacci sequence, $2\times  \partial n_{x_0} = \partial N_{+x_0}+ \partial D_{-x_0}$, and thus asymptotically assuming half-golden background-to-effect ratios of observation,      


\begin{equation*} 
\setlength{\abovedisplayskip}{0pt} 
\setlength{\belowdisplayskip}{0pt} 
        \noindent 
\renewcommand{\arraystretch}{0.1}
          \begin{tabularx}{\linewidth}{>{\hsize=0.9\hsize\linewidth=\hsize}X 
>{\hsize=0.1\hsize\linewidth=\hsize}X} 
           \begin{equation}  \label{eq-intro-golden}
   \frac{\partial N_{+x_0}}{\partial D_{-x_0}} \rightarrow \frac{\phi}{2}. 
\end{equation} &  
  \begin{equation*} 
(m \gg 1)
\end{equation*}
\end{tabularx} 
    \end{equation*}


In conclusion, these equations describe systems that permute effect observations, but whose number of \textbf{effect observations are limited to under-factorial sample sizes}. Samples following half-golden background-to-effect ratios, Eq.(\ref{eq-intro-golden}), observe effects across approximately the same number of backgrounds across all its populations, leading to effect estimates that have, simultaneously, increasing EV and small CF sustained throughout growth. Fig.\ref{fig-intro}(a,b) summarize these three alternative sample size growth rates.    

The two previous shuffling schemes lead then to two strategies for sample growth which can provide some guarantees for either the generalizability, or generalizability and unconfoundness of effect observations. We dub these EV and EV-CF sample growth patterns when studying samples over increasing spatial scales. The cycle-and-permute scheme was studied in ~\citep{ribeiro-ev} and its sample size requirements correspond to the previous golden rates, Eq.(\ref{eq-intro-golden}). Fig.\ref{fig-intro}(a) illustrates that a 'square' contain all in-sample backgrounds (each a Pascal's triangle) for a fixed factor $a$, and thus $2\times 2^{m-1}$ unique effect observations of $a$. Taking columns to mark time progression, the diagram main diagonal marks the point of insertion of factor $a$ (e.g., insert $a$ in populations $\{d\}, \{d,c\}, \{d,c,b\}$). Its lower triangle records the populations without $a$ (with size $N_{-a}$) and the upper triangle with $a$ (with size $N_{+a}$) (\textit{Sect. Sample Limits on Counterfactual Observations}). 

A key result following from the previous is that, in fully-observed samples, a single square is enough for non-parametric effect estimation (plus an irreducible sampling error) ~\citep{ribeiro-ev}. Fig.\ref{fig-intro}(e) illustrates the expected relationship between maximal accuracy of samples, $\textrm{ACC}\Big( S; X \Big[\, x_0, \; d_{x_0}\,\Big]\Big)$, and sample sizes. The diagram shows sample size divided by square size vs. $ACC$. In complete systems (top-curve), the observation of one square (diagonal) is enough to generate accurate effect observations. In incomplete systems (bottom-curve), the number of effect observations necessary for accuracy can grow factorially with the number of unobserved factors - requiring very large samples to achieve similar levels of accuracy. This is a conservative estimate which can be abated by increases in periodicity and independence in out-of-sample factors, but, in real systems, where factors are typically highly correlated, it seems to be a reasonable upperbound for accuracy (\textit{Sect. Results}).


%


\section{Other Related Work}


The Shapley-value ~\citep{Roth:1988wt} has become an essential tool across disciplines to estimate the importance of variables from the output of black-box systems (i.e., whose inputs can be manipulated exhaustively at will)~\citep{10.5555/3295222.3295230,Boer:2020aa,10.1613/jair.1.12228}. The value can be interpreted as the enumeration of all counterfactual effect observations in a \textbf{fully-observed} system  ~\citep{ribeiro-ev}. This makes the Shapley value an instance of an {U-Statistic} and a permutation-based statistic ~\citep{Lee:1990aa,Hoeffding:1948aa}.  The value $\varphi(a)$ was devised first to quantify the importance of a given player $a$ in a $m$-player game, and it can be written as 

\begin{equation} 
\varphi(a)= \frac{1}{(m-1)!}\sum_{\pi \in \Pi(X-\{a\})}\left [ y(P_a^\pi \cup \left \{ a \right \}) - y(P_a^\pi) \right ],
\end{equation}

where $\Pi(X-\{a\})$ enumerate all permutations $\pi$ of a set of size $m-1$, $y$ is a game utility measure, and $P_a^\pi$ is a possible coalition\footnote{a player set describing a possible cooperation structure in the game with value $y(P^\pi)$ when formed.} among players (not including $a$), formed in the order $\pi$. Each quantity under bracket is a counterfactual observation of the effect of $a$ (i.e., under all distinct backgrounds and their orderings). Eq.(\ref{eq-vol}) counted the number of such observations for each population in a sample. The value is an ideal, as its calculation is NP-complete ~\citep{VdBJAIR21} and, when quantifying variable importance, it assumes there are no unobserved causes in the sample. Due to sample correlations this equation cannot be used, as well, with random sampling. The previous squares were devised to define the concept of 'observed permutations', with which the value can be calculated without assumptions of independent and identically distributed factors ~\citep{ribeiro-ev}. Despite these shortcomings, the Shapley-value formulate crucial relationships among permutation of inputs when calculating sample statistics and, respectively, their unbiasedness and 'fairness' ~\citep{10.1613/jair.1.12228,Boer:2020aa}. Calculating the \textbf{expected number of permutations that can be enumerated in samples, or locations, as proposed here, should thus sets strict bounds for their unbiasedness}, and offer a finer-grained illustration of these relationships. While the relationship between the Shapley-value and fairness of black-box predictors is known ~\citep{Roth:1988wt,10.5555/3295222.3295230}, their relation to generalization is perhaps more surprising ~\citep{ribeiro-ev}. A quantity that becomes central to the formulation of accuracy bounds, Eq.(\ref{eq-problem}), and tradeoffs between the two previous learning problems, is the growth rate, $\omega$, in enumerable permutations across systems' spatial levels. Because the Shapley value cannot be calculated in practice, random sampling is often employed as an approximation in Shapley-based importance ranking ~\citep{10.5555/3295222.3295230}. The type of randomization employed in these systems can be seen mathematically as analogous to GSR shuffling (\textit{Sect. Combinatorial Shuffling}). Using square sampling, instead, is advantageous not only for samples with factor correlations, but, particularly, incomplete samples, where, as formulated, assumptions of large random sampling become unrealistic due to their factorial requirements on effect observations. These gains were demonstrated in ~\citep{ribeiro-ev} and are revisited in \textit{Sect. Results}. Furthermore, the theoretic relationships here elucidate sampling size requirements and EV-CF tradeoffs and limits for non-parametric variable importance and effect estimation.  

A key element of the previous solution is that sample units in the same location share a large number of external and uncontrolled factors. Studying samples with increasingly inclusive types of counterfactual effect observations can reveal conditions for generalization across samples and space. Consider the single factor case first. Let $X{-}\{a\}$ be the set of external factors for population $\{a\}$.  In a random sample with a single treatment indicator $a$, it follows that $\Expect \{\;p[a | \, \boldsymbol{\mathcal{D}}(X{-}\{a\})] \;\}= 2^{-1}$, as, at each 2 time intervals, we are expected to rebalance (random variables are bold). This is the rationale underlying, for example, Randomized Control Trials ~\citep{Montgomery:2001aa}. A location with this property has a single balanced populations, $\{a\}$, and common external factors, $X{-}\{a\}$. We can alternatively say that $p(a\;|\; \boldsymbol{\mathcal{D}}(X-\{a\})) = 0.5$, or, $a \perp (X{-}\{a\}) \; | \; \boldsymbol{\mathcal{D}}(X{-}\{a\})$, which are typical non-confounding conditions ~\citep{Rubin:2005aa,Reichenbach:1956aa}. If units in the location share the same external factors, and have the same number of members with $a$ as without $a$, then expected outcome differences between them correspond to $a$'s effect, conditional on the common variation, $\Expect [\Delta y(a) \; | \; x_0 = \boldsymbol{\mathcal{D}}(X{-}\{a\})] = y(x_0{+}\{a\}) - y(x_0{-}\{a\})$. Learning systems and agents in such locations operate with fair estimates of $a$'s impact (albeit, with low EV). In a square, in contrast, \textbf{all $\boldsymbol{m}$ sample factors are balanced simultaneously} ($m>1$). While single-factor balance requires a binomial series, balancing several requires Fibonacci  - i.e., square 'altitude' expansion (\textit{Sect. Spatial Sample EV-CF Growth Patterns}). Each population, in this case, follow asymptotic sample size rates $\Expect [\frac{\partial N_{+a}}{\partial N_{-{a}}}] = \frac{\phi}{2}$. Square accumulation thus increases the EV of all its populations simultaneously ~\citep{ribeiro-ev} - making it easier, for example, to understand limits in sample accuracy across scales, Eq.(\ref{eq-problem}). 

\section{A Combinatorial Perspective on Sample Growth} 

\begin{figure} 
\centering 
\includegraphics[width=0.75\linewidth]{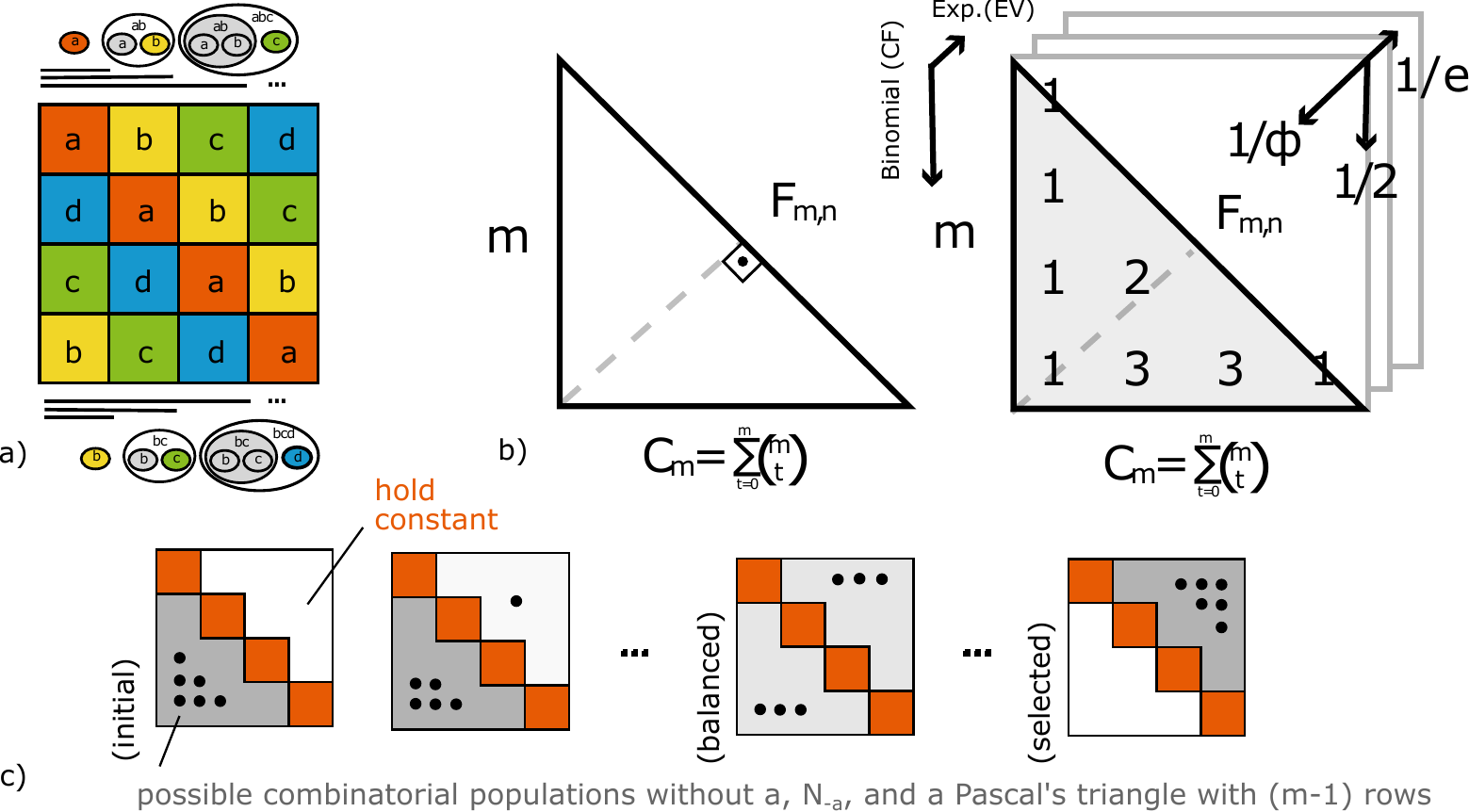}\\ 
 \caption{\footnotesize \textbf{(a)} a $m{\times}m$ Latin-Square ('square') for a sample unit $x_0$ ($m=4$) and the combinatorial relationships among the sample units placed across its different cells (Venn diagram, factor intersections are shown in grey and singleton differences in color), \textbf{(c)} Binomial ($\frac{1}{2}$), Fibonacci ($\frac{1}{\phi}$), and Exponential ($\frac{1}{e}$) rates across squares lead to hyperbolic relations among population sizes, each square's triangle altitude (dashed) is related to samples' background enumeration rate, $\omega$, \textbf{(c)} a sample population 'sweep' for factor $a$, where the rate of insertion of $a$ in populations is held constant, the figure illustrate population sizes as dots and three phases of sample growth: initial (no population has $a$), balanced (the same number of populations have and don't have factor $a$), and possibly selected (where all populations have $a$).}\label{fig-intro2} 
\end{figure}

\subsection{Sample Limits on Counterfactual Observations} 

Statistical Sample Growth can be seen as the enumeration of populations, and their counterfactual observations.  Starting with a population $x_0$, each possible sample growth trajectory (e.g., $x_0, x_1, x_2,...$) is an incremental, and temporally or spatially ordered, observation of the impact on $y$ of gaining, or losing, a set of factors, from the $m$ possible.  In other words, each step in this trajectory is a counterfactual effect observation for $x_0$, $\Delta y(x_0-x_1) = y(x_0) -y(x_1)$ (where the former difference is over sets and the latter over scalars). A counterfactual effect observation is thus defined by: (1) a difference in factors (what changed, $x_0-x_1 \subseteq X_m$), an intersection of factors (what remained the same, $x_0\cap x_1 \subseteq X_m$), and finally, a difference in outcomes (the observed scalar 'effect', $\Delta y(x_0-x_1)\in \mathbb{R}$). There are therefore $2^{m-1}\times (m{-}1)!$ possible counterfactual effect backgrounds for a sample with dimension $m$. Unsurprisingly, this number coincide at the limit with the number of unique shuffles of $m$, established with group-theoretic arguments ~\citep{Diaconis:1983uo}. For each trajectory and time, we can consider the biasedness and generalizability afforded by the accumulated samples to, for example, black-box predictors of $y$, and whether their performance is related to the increasing set of accumulated counterfactuals. Counterfactual reasoning is central to Artificial Intelligence.  For example, most game-theoretic solutions, such as the Nash equilibrium, are formulated from counterfactuals  (i.e., what would happen to a player's utility if it took a given action, but all else remained constant).  Understanding the statistical properties of counterfactual observations ex-post (i.e., their generalizability, biasedness), or sets of such observations, should therefore be central to the design and analysis of AI, multi-agent and learning systems. 

The set of all counterfactuals accumulated by sample growth at one instant can be visualized with a Latin-Square ('square'), Fig.\ref{fig-intro2}(a). The square will serve as basis for non-parametric effect estimates across sample factors. For a fixed unit or population $x_0$, it represents a stratification, or 'placement', of all \emph{other} populations, $x_i$, across square cells, with repetition. The completeness or incompleteness of squares, for each  $x_0$, will have a stipulated impact on the EV or CF of their effect observations. In particular, for $m$ factors ($a$, $b$, $c$,..., $[m]$) the square enumerates all singleton effect observations possible from a sample's $m$-way effect differences. Its first column contains counterfactual effect observations for $\{a, b, c, ..., [m]\}$ (i.e., conditioned on all other $m-1$ factors being the same as $x_0$). The second column contains singleton effect observations possible from the previous observations (with size $1$ difference and $m-2$ intersection with $x_0$). These effect observations are thus conditioned on one further factor observation (i.e., on the factor in the preceding column). The third column contains singleton effect observations possible from the previous observations (size $1$ difference and $m-3$ intersection). Fig.\ref{fig-intro2}(a) illustrates these combinatorial patterns with Venn diagrams for each cell, where a cell's pairwise intersecting factors are grey and singleton differences are colored. This iterative procedure enumerates all possible singleton effect observations in a sample.  The square diagram shows only the singleton effects (cells), with their effect conditioning factors indicated by the preceding factors in the diagram and other effects implied. Each of its diagonals contains all observations for the effect of a fixed factor in the sample. The square of order $m \times m$, as a whole, contain effect observations where all factors are observed under all $m$-cycles of a fixed permutation (e.g., $\{a, b, c, d\}$ in Fig.\ref{fig-intro2}(a)). Squares of increasing orders thus captures effect observations under increasing Markovian orders (i.e., conditioned across larger times or backgrounds).  The relationship of sample permutations to measurements' unbiasedness is a cornerstone of the most widely-accepted Theory of Non-parametric Statistics, U-Statistics ~\citep{Lee:1990aa,Hoeffding:1948aa} and of Shapley value based estimates of black-box predictive performance (\textit{Sect. Other Related Work}). The relationship to  generalizability has been discussed in ~\citep{ribeiro-ev}, and is reviewed, and expanded, below.    

The full set of $\sum_{t=0}^m \binom{m}{t} = 2^m$ effect observations observable in a sample of dimension $m$ collect $1$ square for each of its sample populations, $x_i \subseteq X_m$. It suggests then a natural sample limit for the generalization of effects. In large scale spatial samples, like the studied below, these strict sample limits become very clear. We can, as result, visualize the lower triangle of the square as the $m-1$ Pascal's triangle of size $m-1$, containing the $2^{m-1}$ unique sample populations, $x_i \in [0,1]^{m-1}$, without the factor in the square's main diagonal. Fig.\ref{fig-intro}(b) exemplify the resulting phases of samples with factor $a$ under growth: no unit includes $a$ (initial), as many units include $a$ as not (balanced), and all include $a$ (selected). Eq.(\ref{eq-intro-hyperbolic}) should hold across all such scenarios ($N_{+a},N_{-a}{>}0$).    

\subsection{Combinations, Permutations and Partial Permutations} 

The statistical concept of a 'population' is often associated with combinatorial combinations, as a set of sample units with a given attribute combination (e.g., high-income white males). There are thus $\binom{m}{t} =  \frac{m!}{t!(m-t)!}$ populations of \textit{size} $t$. A problem with this definition is that it leaves unspecified all non-population factors. To \textbf{define a population we imagine, instead, that we fix the $m-t$ population factors and vary (i.e., 'permute') all $t$ non-population (i.e., 'external') factors}. This leads to a combinatorial structure known as a partial permutation.  The number of partial permutations for a population of size $t$ is $\binom{m}{t} \times D_{m-t}$, where $D_{m-t}$ is the number of \emph{derangements} (permutations without overlaps),

\begin{align}
D_{m-t}=\sum_{k=0}^{m-t} \frac{(-1)^k}{k!}.
\end{align}

 The full set of $m!$ permutations of size $m$, and all sample growth trajectories, can be formulated as sets of partial permutations, using a well-known definition for factorials,

\begin{align} 
m! &=\sum_{t=0}^m \binom{m}{t} \times D_{m-t} \label{eq-derangement}\\ 
&=  \Big[ \underbrace{ {\cosh(m{-}1)+\sinh(m{-}1)} }_{ \frac{\cosh(m{-}1)}{\sinh(m{-}1)} \xrightarrow{m} \omega } \Big] \times (m{-}1)!   +1.  \label{eq-derangement-2} 
\end{align} 

The term $C_m{=}\sum_{t=0}^m \binom{m}{t}$ in Eq.(\ref{eq-derangement}) corresponds to a single Pascal triangle and half-square (i.e., one set of all differences) for each sample population ~\citep{ribeiro-ev}, and Eq.(\ref{eq-derangement}) to all squares.  The number of observed permutations in a sample can thus be specified succinctly by its number of squares and their derangements. Samples with no missing variables require the observation of few derangements (no relevant exogeneous variation) for accurate effect observations, while incomplete samples require the observation of many derangements ~\citep{ribeiro-ev}.    The odd and even parts of Taylor's expansion of Eq.(\ref{eq-derangement}) leads to hyperbolic trigonometric functions, Eq.(\ref{eq-derangement-2}) (proof in the \textit{Supplementary Material}). They indicate the 'period' in which full sets of permutations are collected.  The parametric equations for the hyperbola's right branch in $(x,y)$ cartesian coordinates, Eq.(\ref{eq-intro-hyperbolic}),  are $x=\omega \times \cosh(N_{+a})$ and $y=\omega \times \sinh(N_{-a})$, Fig.\ref{fig-intro}(e). We will see that these quantities are related to in-sample effect to background enumeration rates, $\omega$, across time or spatial scales in systems. This quantity will be essential to understanding statistical tradeoffs across growing samples.

\subsection{Spatial Sample EV-CF Growth Patterns} 

A singleton population $\{a\}$, defined as in the previous section, is represented by a square diagonal, Fig.\ref{fig-intro}(b). Behind each square and Shapley-value calculation (\textit{Sect. Other Related Work}) is an experimental procedure: add $a$ to every variation of other populations; with each insertion, observe before-after outcome differences, $\Delta y(a)$. An unbiased effect estimate for $a$ is an average across all possible observations, and constitutes an U-Statistic ~\citep{Lee:1990aa}. There are $F_{m,n}=\sum_{t=0}^{m-1}\binom{m-t}{t}$ such sequential observations\footnote{$\binom{m-t}{t}=0$, when $t>m$. }. To generate all of them, we need to fix each effect observation's first, second, third, etc. factors in order. $F_{m,n}$ correspond to the sum of the number of observations necessary to fix any first factor, $\binom{m-1}{1} = m{-}1$ , then $\binom{m-2}{2}$ to fix a second from the remaining, etc. until all $m-1$ factors are used.  


The relationship in Eq.(\ref{eq-intro-hyperbolic}) corresponds to the Cartesian equation of a rectangular hyperbola, $N_{+x_0} \times N_{-x_0} = c$, where $c$ is constant (although well-known, this is formulated in the \textit{Supplementary Material} for completeness).  According to the previous, \textbf{these two quantities have different limits}, however, $N_{+x_0} \in [1,C_m]$ and $N_{-x_0} \in [1,F_{m,n}]$. The relationship can thus describe In large-populations sample limits by substituting $N_{+x_0} = C_m$ and in $N_{-x_0} = F_{m,n}$ in Eq.(\ref{eq-intro-hyperbolic}). As formulated next, the same result can be established from known rates across Pascal's triangle.


The two previous quantities, $C_m$ and $F_{m,n}$, appear in Pascal's triangle (adjacent side and altitude), Fig.\ref{fig-intro2}(b). Since the main diagonal marks $a$'s possible 'times-of-insertion', the square's upper triangle contains the set of all counterfactuals with $a$, and the lower, without $a$, Fig.\ref{fig-intro2}(b).  In respect to effect observations, we say that each individual effect observation is observed under $F_{m,n}$ in-sample backgrounds for each derangement $D_n$ (or twice this value in balanced samples). The sample background enumeration rate $\omega$, at time $t$, is thus defined as $\omega = \frac{F_{m,n}}{D_n}(t)$, or,  the number of in-sample background observations, $F_{m,n}(t)$, per derangement, $D_n(t)$, across all populations in the sample. 

The growth of $C_m$, $D_n$ and $F_{m,n}$ for $N_{+a}$ or $N_{-a}$ assume Pythagorean relations\footnote{the equation uses the Pythagorean theorem in its reciprocal form, as it includes the triangle's altitude.}, Fig.\ref{fig-intro2}(b),

\begin{equation*} 
\setlength{\abovedisplayskip}{0pt} 
\setlength{\belowdisplayskip}{0pt} 
        \noindent 
\renewcommand{\arraystretch}{0.1}
          \begin{tabularx}{\linewidth}{>{\hsize=0.4125\hsize\linewidth=\hsize}X 
>{\hsize=0.4375\hsize\linewidth=\hsize}X>{\hsize=0.15\hsize\linewidth=\hsize}X} 
    \multicolumn{2}{X}{         \begin{equation} 
  \label{eq-pythagoras-1} 
  \Big(\frac{\partial C_m}{\partial D_n}\Big)^{-2} + \Big(\frac{1}{\partial D_n}\Big)^{-2} = \Big(\frac{\partial F_{m,n}}{\partial D_n}\Big)^{-2}, 
\end{equation} }&  
  \begin{equation*} 
\Big(\frac{\partial C_m}{\partial F_{m,n}}\textit{ const.}\Big) 
\end{equation*}
\end{tabularx} 
    \end{equation*} 

Eq.(\ref{eq-pythagoras-1}) suggests the visualization of sample growth as hyperbolae\footnote{the equation for a hyperbole is $(\frac{x}{a})^2-(\frac{y}{b})^2 = r$, with $a$ and $b$ its vertices and $r$ radius.} with increasing radius $D_n$, Fig.\ref{fig-intro}(e). In this limiting expression of Eq.(\ref{eq-intro-hyperbolic}), $C_m$ corresponds to all possible individual sample populations, $N_{+x_0}$ and $N_{-x_0}$, and $F_m$ to in-sample effect observation backgrounds. The figure shows the hyperbolic asymptotes $C_m=F_{m,n}$ and $C_m=-F_{m,n}$ (dashed). They represent growth with constant EV, $\partial D_n =0$. The figure also shows the asymptotic sample (vertical black line) where exactly all observations have factor $a$, $F_{m,n}=0$. \textbf{Under this condition, no estimator, algorithm, or agent is able to estimate $\boldsymbol{a}$'s effect}. It represents the sample with minimum EV, while outward hyperbolae, samples with increasing EV. Growth in this direction follow a Fibonacci series, whose rate is the Golden number. It is well known that the rows, columns and diagonal of Pascal's triangle are associated with binomial, exponential and fibbonacian rates. Notice then that $\frac{\partial D_n}{\partial C_m} \in [1/e,1/2]$, as growth can range between $\frac{\partial C_m}{\partial m}=2$, and $\frac{\partial D_n}{\partial n}=1/e$, Fig.\ref{fig-intro2}(b). The first is due to $C_m = 2^m$, and the second was famously established by Euler ~\citep{SandiferCharlesEdward2007HEdi}. In the previous nomenclature, the first is associated with balanced or Unconfounded (CF) growth, and the second with EV sample growth. The golden ratio is associated, in contrast, with high-dimensional balanced growth of samples and populations, EV-CF growth, and with squares, Eq.(\ref{eq-pythagoras-1}). 

More specifically, squares are associated with the assumption that $\frac{\partial C_m}{\partial F_{m,n}}$ is constant across factors (i.e., hyperbolae with constant radius), Eq.(\ref{eq-pythagoras-1}). It indicates that factors' diagonals are the same size, and the population structure is, overall, a 'square'. The following are known hyperbolic relationships,

\begin{equation*} 
\setlength{\abovedisplayskip}{0pt} 
\setlength{\belowdisplayskip}{0pt} 
        \noindent 
\renewcommand{\arraystretch}{0.1}
         \begin{tabularx}{\linewidth}{>{\hsize=0.4125\hsize\linewidth=\hsize}X 
>{\hsize=0.4375\hsize\linewidth=\hsize}X>{\hsize=0.15\hsize\linewidth=\hsize}X} 
    \multicolumn{2}{X}{         \begin{equation} 
  \label{eq-pythagoras-2} 
   tanh(n)=\frac{\sinh(n)}{\cosh(n)}= {\sqrt{ 1 + \Big(\frac{\partial D_n}{\partial C_m}\Big)^2}}=\omega^{-1}, 
\end{equation}
}& 
  \begin{equation*} 
 \Big(\frac{\partial F_{m,n}}{\partial D_n} = \omega\Big) 
\end{equation*}  
\end{tabularx} 
    \end{equation*}  

These equations suggest expressing sample background enumeration rates $\omega$ in terms of $\tanh(n)$\footnote{with $n =\arctanh(\omega^{-1}) = \arctanh(\frac{D_n}{F_{m,n}})$, which, lets $n$ be the number of accumulated derangements per fixed $F_{m,n}$ (i.e., per square), as expected. }.  Also note that this definition for $\omega$ coincides with that of Lorentz factor $\gamma$ ~\citep{carroll2003spacetime,Hucks:1993aa}, best known as a time correction between frames-of-reference in the physical sciences.  Here, it preserves frequency relations among factors, $N_{+a}/N_{-a}$, under changes of basis of the type $x{=}x_0{+}\{a\}$ and $x{=}x_0{-}\{a\}$. As suggested by Borel ~\citep{Borel:1914aa}, it is natural to think of the transformation as a hyperbolic rotation (analogously to the typical trigonometric). We will illustrate many of these mathematical abstractions using real-world spatial data in \textit{Sect. Results}.

\section{Results}

\begin{figure} 
\thisfloatpagestyle{empty} 
\centering 
\includegraphics[width=0.75\linewidth]{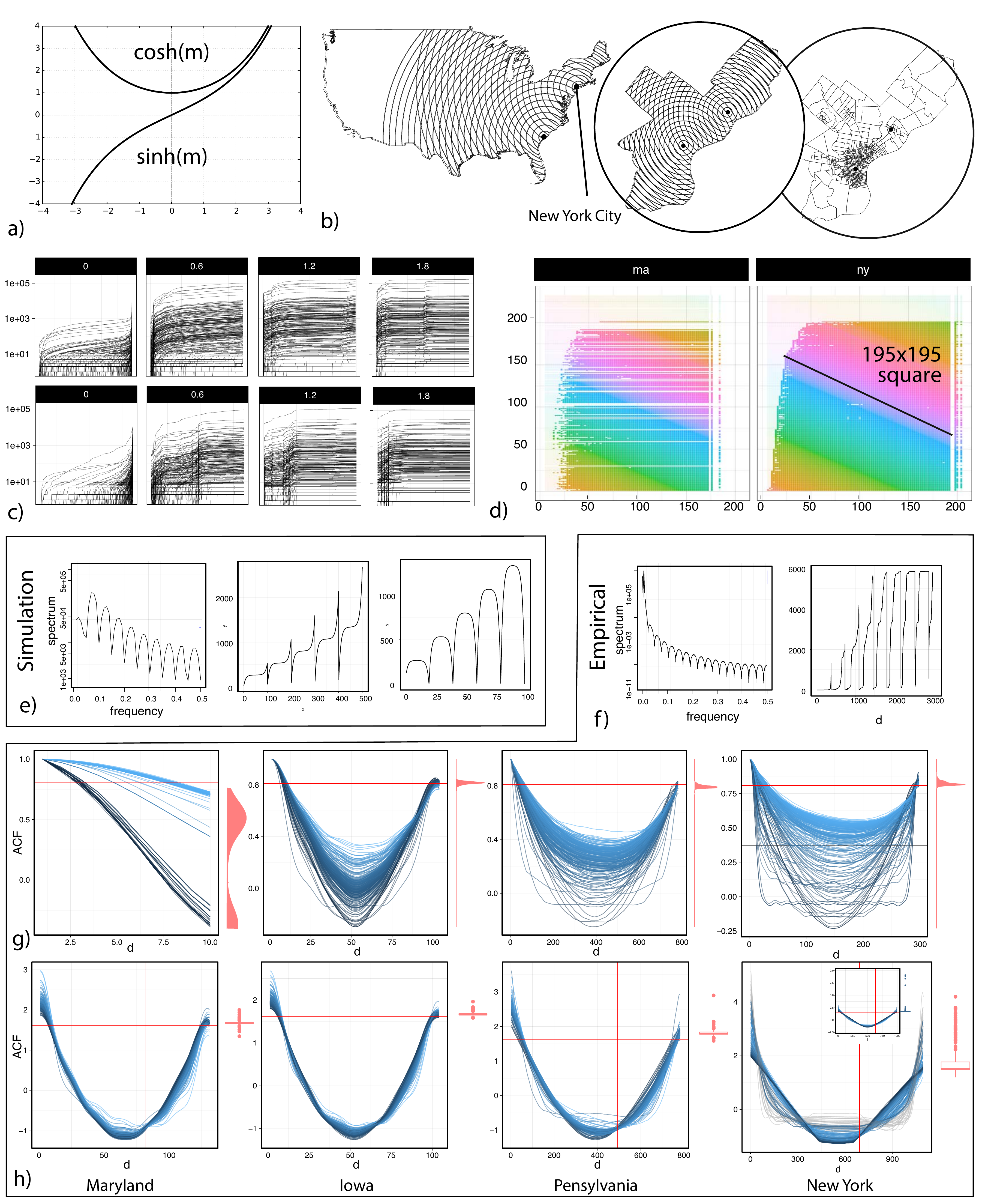}\\ 
\caption{\footnotesize \textbf{(a)} $\sinh$ and $\cosh$ functions, \textbf{(b)} increasing spatial-levels at two example locations (national and city-levels), (rightmost panel) finest spatial-level for New-York City,  \textbf{(c)} occupation frequency ranks vs. location across 4 example scales, each curve is an occupation, \textbf{(d)} enumerated Latin-Squares histograms for Massachusetts and New York, the latter has a square with almost all occupations, \textbf{(e)} periodogram of $cosh(100){+}sinh(100)$, $\sinh(100)$ and $\cosh(100)$, \textbf{(f)} per-occupation periodogram and series example from (c), \textbf{(g)} auto-correlation vs. spatial-level trace catenaries (free-hanging ropes) for each occupation (1880), probability distribution of their slack (red, sidepanel) indicate a fixed $\omega$ per factor at $0.81$ (red horizontal line), Eq.(\ref{eq-slack}), \textbf{(h)} standardized catenaries across \textit{all} years, boxplots (red, sidepanel) show slack invariance and constant ratio between $\sinh$ and $\cosh$ growth for all locations, years and occupations,  $m\times(1-\frac{1}{e})$ (red vertical line) is a fixed point in binomial-exponential (EV-CF) to exponential (EV) rate transitions.}\label{fig-res1} 
\end{figure} 

We will now illustrate the formulated combinatorial and statistical generalizability limits in an important real-world problem: out-of-sample economic growth prediction across increasing spatial extensions (i.e., samples with increasing census individuals). Data used encompasses microdata of American decennial censuses from 1840 to 1940, and approximately 65 billion individual-level records. This time range corresponds to the decades of American urbanization. We consider the economic and demographic changes as we go, spatially, from the household spatial-level, $d_0$ in lat-lon distances, all the way to the national level, for each studied year. We thus create samples with units at arithmetically increasing levels, $d_{t+1}{=}d_t{+}\Delta d$ (starting from $d_0$). We repeat this for approximately 60K American locations, $x_0$. Each full spatial analysis is then reproduced independently across years (avoiding issues related to extended longitudinal data). Fig.\ref{fig-res1}(b, middle) shows two locations in New York City, which share a large amount of external variation (i.e., economic and demographic variations across the rest of the country). The resulting nation-wide transversal captures  combinatorial patterns of populations' differences and overlaps in samples, for all $x_0$, as we increase scale.  Our main goal is to illustrate how, consequently, \textbf{generalizability change across spatial-scales}, according to the stipulated model and limits. We first consider sample correlations and sizes, demonstrating they follow the previous hyperbolic relationships. We then repeat previous out-of-sample prediction tasks with this new census data and increasing spatial levels - thus adding to previous evidence presented to a combinatorial counterfactual model for sample generalizability ~\citep{ribeiro-ev}  . 

\subsection{Descriptive Statistics (Sample Sizes and Correlations)} 

We illustrate the consequences of Eq.(\ref{eq-pythagoras-2}) to sample properties using Autocorrelation functions (correlations) and hyperbolic co-tangent ($\coth$) regressions (sample sizes) in large-scale census data. These considerations will be key to solving our main problem, Eq.(\ref{eq-problem}), as the accuracy of agents and algorithms operating in samples are directly determined by sample sizes and their combinatorial patterns ~\citep{ribeiro-ev}. Economic distribution across space can be described by the primary occupation and industry of all census individuals ~\citep{Balland:2020aa,Inho:wo} (e.g., 'carpenter' or 'executive assistant'). We start with this set of variables, and discuss the full set of variables, including non-economic, in the next section. Fig.\ref{fig-res1}(c) illustrates empirical frequencies for all occupations (each a curve) at 4 different spatial-levels in Massachusetts (MA) and New York (NY), 1880. They were the country's economic centers until the 19th century. The distribution has the familiar shape of a wave that moves to the left. New York reaches a stationary shape at a lower level $d_{sq}$. We demonstrate these correspond to levels where squares are completed across factors. All squares in a location can be enumerated through an expensive computational procedure ~\citep{ribeiro-ev}. Fig.\ref{fig-res1}(d) shows histograms, where each color corresponds to one of $220$ occupations. NY has a spatial square that extends to almost all occupations, while MA has missing occupations (horizontal gaps) in comparison.  

\subsubsection{A 'Hanging-Rope' Model for Unbiased Sample Growth} 

The Catenary is a curve with long scientific history. Unlike circles and geodesics, they are sums of exponentials. Catenaries describe a free-hanging rope ~\citep{Cella:1999aa}. Their equation in $(x,y)$ Cartesian coordinates is $y = \cosh \left(x\right)$, and their length is $l = \sinh(x)$, making them useful to demonstrate the previous model, Eq.(\ref{eq-pythagoras-2}), and increases in enumerable permutations across spatial levels. We demonstrate that both {the observed shape, Fig.\ref{fig-res1}(h), and parameters, Fig.\ref{fig-res1}(h, boxplots), of spatial correlations follow predictions from the previous model}.  Before considering catenaries, however, Fig.\ref{fig-res1}(a,e) illustrate the overall shape of the previous hyperbolic functions, and their frequency-based representations. Fig.\ref{fig-res1}(a) depicts $\cosh(n)$ and $\sinh(n)$, and, Fig.\ref{fig-res1}(e) the periodogram of $cosh(n){+}sinh(n)$, $\sinh(n)$ and $\cosh(n)$ where $n=100$. Fig.\ref{fig-res1}(f) illustrates the \textit{empirical} periodogram of curves in Fig.\ref{fig-res1}(c), which resemble the simulated. 

Fig.\ref{fig-res1}(g) shows auto-correlations (ACF) for all locations across 5K spatial-levels (as those illustrated in Fig.\ref{fig-res1}(c)), until the state level. They trace catenaries.  The horizontal line $y= 1.0$, of unitary correlation, is associated with the limit $F_{m,n}=0$ where, despite the increasing scale, no population differences are added to the increasing samples. Each single catenary is a set of samples with constant $C_m/F_{m,n}$, which is a \textbf{defining property of squares}, Eq.(\ref{eq-pythagoras-1}). Fig.\ref{fig-res1}(g) illustrates 4 typical cases among states. Plots for all states are available in the Supplementary Material. Maryland has linear decreases in auto-correlation. From 1840, the USA economy and cities become increasingly interdependent. After 1900, no longer any state had such linear correlation signatures. Periodic and linear (zig-zag) auto-correlations, with period $m/2$, are related to non-increasing EV, Fig.\ref{fig-intro}(e, black vertical line). Periodic and exponential correlations, without growth, correspond to catenaries with $h=0$ (where a system returns to its original state after a lag). The defining characteristic of the catenary is $\nicefrac{\partial y}{\partial x} = \nicefrac{l}{h}$, where $l$ is its length and $h$ its 'slack', or, difference in height, $y$, between its two hanging points. Standardizing catenaries ~\citep{Cella:1999aa} (i.e., making $l$ unitary and $h$ constant)\footnote{$\nicefrac{\partial y}{\partial x} = \nicefrac{\partial \cosh (x) }{\partial \sinh (x)}= \nicefrac{\sinh (x) }{\cosh (x)} = \tanh(x)$.} thus makes its slack $h$ indicate $\omega$ during sample growth,

\begin{equation*} 
\setlength{\abovedisplayskip}{0pt} 
\setlength{\belowdisplayskip}{0pt} 
        \noindent 
\renewcommand{\arraystretch}{0.1}
          \begin{tabularx}{\linewidth}{>{\hsize=0.8\hsize\linewidth=\hsize}X 
>{\hsize=0.2\hsize\linewidth=\hsize}X} 
           \begin{equation}  \label{eq-slack}
   h = \omega \approx \frac{\phi}{2},  
\end{equation} &  
  \begin{equation*} 
\Big(\begin{matrix}\partial D_{-x_0} \approx 1, \\[0.25cm] m\gg1 \end{matrix} \Big)
\end{equation*}
\end{tabularx} 
    \end{equation*}


which, according to Eq.(\ref{eq-intro-golden}), should assume half-golden values for small $D_{-x_0}$.  Fig.\ref{fig-res1}(h) shows standardized catenaries for all years and locations. It indicates that $\tanh$ per factor remains constant across a range of levels, up to $d_{sq}$, starting at the local. This was anticipated by Eq.(\ref{eq-pythagoras-2}). The rate, up to $d_{sq}$, is $81\%$ of correlation. Fig.\ref{fig-res1}(h, sidepanel, red) shows \textbf{box-plots} for $h$, across all levels, years, occupations, and locations. For all spatial-levels below $d_{sq}$, factors remain balanced, with binary-exponential rates (i.e., hyperbolic functions with period $m/2$). Levels above $d_{sq}$ reverse to exponential growth. We called this a transition between EV-CF and EV growth. This is indicated in plots by the dislocation of the catenary center from $m/2$ to $m(1-1/e)$ (red vertical lines). We reproduce the same results, Fig.\ref{fig-res1}(g,h), with standard \textbf{Pareto regressions} in \textit{Sect. Methods} of the \textit{Supplementary Material} - as alternative to these graphical depictions. Fig.\ref{fig-res2}(d) shows estimated levels $d_{sq}$ for all states, across years. Levels $d_{sq}$ converge across years for all states. New York has 2-level squares. Fig.\ref{fig-res1}(h, upright-panel) shows catenaries for its lower-level square, and Fig.\ref{fig-res2}(b) illustrate levels cartographically. The two squares' factors are disjoint (gray, lower factors taking exponential rates in higher). American states have had through their histories very different work forces and regional distributions. While catenary lengths are different across occupations, Fig.\ref{fig-res1}(g), their slack (and $\cosh$-$\sinh$ growth ratio) \textbf{remains invariant across all locations, years and occupations}, Fig.\ref{fig-res1}(h). According to previous discussions, these plots thus illustrate graphically combinatorial constraints for the unbiasedness and predictiveness of learning systems across space.  We return to this discussion in \textit{Sect. Predictive Statistics}. 

\subsubsection{Permutations in Heterogeneous Samples} 

\begin{figure} 
\centering 
\includegraphics[width=0.65\linewidth]{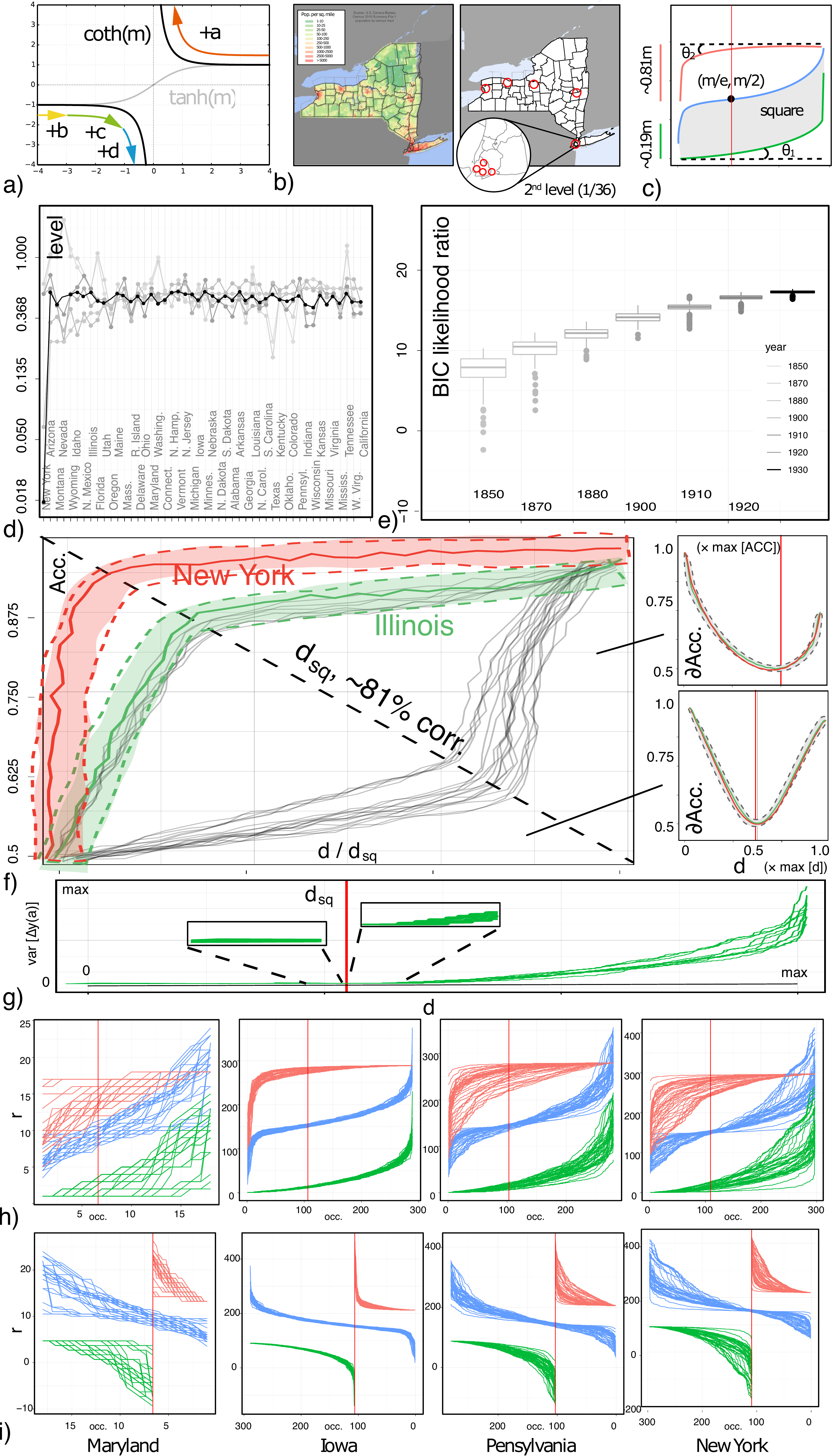}\\ 
\caption{\footnotesize \textbf{(a)} $\coth$ and $\tanh$ functions, colored arrows illustrate square factor frequency increases with scale, \textbf{(b)} New York (NY) state population density (left), NY has squares at two levels $d_{sq}$, at $0.018$ and $0.65$ lat-lon distances (red circles), \textbf{(c)} schematic depiction of frequency rank vs. spatial-scale plots in (h),  \textbf{(d)} $d_{sq}$ (spatial-level of first square) for states and years, \textbf{(e)} BIC likelihood of $\coth$ over a Zipf model, \textbf{(f)} state-of-the-art growth prediction's accuracy with increasing spatial-level $d$, $d_{sq}$ is diagonal (dashed),  \textbf{(g)} min. (green) and max. (red) frequency ranks across locations, each curve is a scale, blue curves indicate square size, which follows a $\tanh$ function, \textbf{(h)} $\coth$ model, as illustrated in (a), with empirical data.}\label{fig-res2} 
\end{figure} 

 Zipf's law and distribution are central to the study of city size distributions ~\citep{Newman:2005aa,Gabaix:1999aa,Berliant:2018wj}. The law is based on a frequency ranking of studied factors, and thus, on one of their permutations. It is, here, associated with homogeneous samples (i.e., samples with little across-factor variation). Fig.\ref{fig-res2}(a) depicts the overall shapes of $\tanh(n)$ and $\coth(n)$ functions. Fig.\ref{fig-res2}(h) shows occupations' minimum frequency rank, $r_{0}$ (green), across all locations in increasing spatial levels, as well as their maximum rank, $r_{\omega}$ (red). The former is the minimum frequency ranking order of one given occupation across all the level's locations. The latter is the maximum (these are formulated explicitly in \textit{Sect. Methods}). The latter is related to Zipf's frequency rankings and the Pareto distribution (\textit{Sect. Methods}), as the three are Power-laws. Each curve in the figure corresponds to one spatial-level and occupation. With a homogeneous sample, we expect one highest-rank industry across all locations, and thus $r_{\omega}-r_{0} = 1$. What we observe, however, is that factors are ranked in constant-sized ranges, as visualized in squares. Each factor is the highest ranked in \emph{some} location, the second in other, etc. These rankings define an arithmetic series $-$ $r_{0}, r_{0}+1, r_{0}+2, ..., r_{\omega}$ $-$ for each factor. The series has mean $\bar{r}=\frac{r_0 + r_\omega}{2}$, which is also shown (blue). The previous model predicts both that $r_{\omega}-r_{0}$ is constant, and that it reflects the enumeration rate $\omega$. Fig.\ref{fig-res2}(h) shows that empirical rankings have constant $r_{\omega}-r_{0}$, with increasing $r_{0}$.  A closer examination of both branches (red and green) reveals they correspond to the positive and negative sections of the $\coth(n) = 1/ \tanh(n)$ function, Fig.\ref{fig-res2}(i,a). 

Imagine the following process: pick a location $x_0$, and its most and least-frequent factors (i.e., with rank $1$ and $m$). Label them, respectively, $a$ and $z$. Balance $z$ to match $a$'s frequency. Move one spatial-level up, pick another $z$, balance, and repeat. This is the process described by Eq.(\ref{eq-intro-hyperbolic}).  Each square row corresponds to a single derangement and background, $\partial n_{+a} = \omega \times \partial D_a$ is the number of units in cell $a$. The cost to balance each $z$ is thus $n_{+a}$/$\omega$ per row.  For all locations $x_0$, and levels $d_0 \leq d \leq d_{sq}$,

  \begin{equation} 
\begin{split} 
 n_{+a}  \times \frac{1}{\omega}- n_{+z \in X-\{a\}}  &= 0,\\ 
n_{+a}  -   n_{+z \in X-\{a\}} \times \coth(n) &= 0.\\ 
\end{split} 
\end{equation} 

  The $\coth$ function has the interesting property of \textbf{separating, by sign, each location's 'background' and 'effect' phases}, and describe more directly how squares are completed. This is illustrated in Fig.\ref{fig-res2}(a) as one hyperbolic rotation, with subsequent square derangements leading to others. 
  
  Methodologically, this suggests fitting a $\coth$ function to observed frequency ranks. A Zipf-distribution can be fit by Power-law or Pareto distribution regressions (\textit{Sect. Methods}).  Enumeration rate increases imply increasing permutations - and thus differences between min. and max-frequency ranks. This predicts that Zipf-Pareto regressions will become increasingly inaccurate (compared to $\coth$), as cities become more heterogeneous. Fig.\ref{fig-res2}(e) shows increase of up to 18 times fit likelihood favoring the $\coth$ model throughout the studied period, according to a Bayesian Information Criterion.   
  
\subsection{Predictive Statistics (Reproduction of ~\citep{ribeiro-ev})} 

What impact does the presence of squares in samples have statistically (in respect to bounds to their predictiveness and biasedness)? This was formulated theoretically, and demonstrated practically in simulated, cohort, experimental, economic, and genetic data ~\citep{ribeiro-ev,ribeiro-periodic}. Fig.\ref{fig-res2}(f) demonstrates a further result, using census microdata, with an Accuracy vs. Spatial-level plot. Samples in the previous section contained sample units' primary occupations ~\citep{Census.-Q-United-States.-Bureau-of-the-Census.:1951aa,Osborne:2005aa}. This led to binary samples of dimension $m=543$ (and, each unit seen as a 543 length binary vector). For Fig.\ref{fig-res2}(f), all variables in the American micro census were, instead, used ~\citep{ipums}. Each census binary variable lead to one field, each categorical variable to as many binary variables as the size of their domain (as defined by the census) and continuous variables to 8-bit vectors (corresponding to their 8 quantiles). The final sample had 10.055 variables, including information on a broad range of population characteristics, including fertility, nuptiality, life-course transitions, immigration, internal migration, labor-force participation, occupational structure, education, ethnicity, and household composition. These variables can be correlated, colinear and spurious. Each of the multiple state-of-the-art classifiers employed next will deal with these statistical pitfalls in their own proposed ways. 

The classification task in Fig.\ref{fig-res2}(f) is to predict whether a given occupation will grow (enlist further members) in the next time interval (10 years ahead), for each location $x_0$. Detailed description of algorithms used, and their hyperparameter optimization, can be found on ~\citep{ribeiro-ev}. They include Neural Network Models, Generalized Linear Models, Boosting Models, Generalized Additive Models, Random Trees, LASSO and Ridge regressions, ANOVA, Support Vector Machine, and stacked meta-learners for all previous algorithms. Spatial levels (and aggregated data) ranged from the local to the national. One million location and year were chosen randomly, each leading to a full set of spatially growing samples.  The figure thus shows the maximum accuracy of 24 state-of-the-art supervised algorithms, to predict whether a given occupation will grow, or not, in a location, as we use data from increasing spatial-levels (starting with the local and reaching all national). Accuracy is defined as the number of accurately classified observations in the held-out sample. Spatial levels $d_{sq}$ for each state are mapped to the diagonal (dashed) in the figure, and each state is a curve. The way accuracy changes across locations largely follow the shape expected by Fig.\ref{fig-intro}(e). Accuracy was averaged across same-state locations to generate these curves. Bootstrap accuracy variation bands (across states' locations) are shown for the two most accurate states, New York and Illinois. We observe that New York gains little from external data, above $d_{sq}$, as it already contains, within its boundaries, high levels of variation. This also implies that, without unobservables, ${\sim}81\%$ of the sample is sufficient for prediction. Homogeneous locations, in contrast, have \textbf{incomplete squares}, and observed predictions are susceptible to external and unobserved variation ~\citep{ribeiro-ev}.  

The right panels in Fig.\ref{fig-res2}(f) shows the \emph{increase} in accuracy, $\nicefrac{\partial ACC}{\partial d}$, of samples encompassing increasing distances (at $0.05$ lat-lon intervals, normalized over their spatial and accuracy ranges) for New York and Illinois. This is the difference in accuracy of models trained at a level $d$ and its predecessor. The top panel shows accuracy of algorithms with samples with observations $d > d_{sq}$, and bottom panel $d\leq d_{sq}$, across all locations in those states (gray ribbons show their standard deviation). These patterned differences in accuracy changes, in the output of supervised black-box algorithms, mirror the shape of functions in Fig.\ref{fig-res1}(g,h). This is expected as the accuracy of systems with characteristics described above increase with pairwise correlations ~\citep{ribeiro-ev}. The functional form for accuracy $\mathcal{F}(d, \mu) = \mu \times \sinh(d)$ take, therefore, hyperbolic forms with distinct parameters\footnote{the same is expected from $\frac{d}{dx}\sinh x = \cosh x$ and results in  \textit{Sect. A 'Hanging-Rope' Model for Unbiased Sample Growth}.}, $\mu = 0.5\times m$ and $\mu = (1-1/e)\times m$, and constant $\tanh$, $0\leq \mathcal{F}(d, \mu)\leq 1$. The 80-20 ratios, observed for correlations in \textit{Sect. A 'Hanging-Rope' Model for Unbiased Sample Growth}, are thus also observed in the outcome of black-box predictors, Fig.\ref{fig-res2}(f). Like before, the functional $\mathcal{F}(d, \mu)$ is an apt description in these systems only because $\tanh$ also remain constant throughout them (i.e., in squares and balanced samples), Eq.(\ref{eq-pythagoras-1}).

Given the balance of samples with $d < d_{sq}$, it is expected for effect estimation to be easier in these samples, in contrast to samples with $d > d_{sq}$.  In ~\citep{ribeiro-ev}, we use multiple simulated scenarios to show that samples having combinatorial conditions like those with $d < d_{sq}$ facilitate causal effect estimation. This is possible because, there, we have ground truth information for the effect of variables. We do not have ground truth in this large real-world system, but the central claim here is that in samples with $d > d_{sq}$ the problem becomes harder. Fig.\ref{fig-res2}(g) shows variance in effect estimation for the popular Shapley-based effect estimation ~\citep{10.5555/3295222.3295230} across all locations $x_0$ and distances $d$, in the previous samples. There is a discontinuity, and significant increase in uncertainty over effect estimates above $d_{sq}$ (and very little below), across all locations. Together, the previous considerations suggest constraints, described by $\mathcal{F}(d, \mu)$,  for supervised prediction and effect estimation in spatial systems. Functions $\mathcal{F}(d, (1{-}1/e)m)$ and $\mathcal{F}(d, 0.5m)$ describe, respectively, patterns of externally-valid (EV) and unconfounded (EV-CF) spatial sample growth.

 \section{Conclusion} 

We studied applications of concepts from non-parametric and counterfactual statistics to sample growth processes; common, for example, in the study of spatial systems. We highlighted sample conditions where $m$ populations can have their effect observations remain unbiased while, at the same time, increasing in generalizability. The set of all squares of size $m$ in a sample is related both to an optimal estimator in Theoretical Statistics and a solution in Cooperative Game-Theory. Hyperbolic functions offered a natural implementation and visualization for these alternative growth patterns.  Increase in generalizability, for $m = 1$, requires exponential sample size growth. Increase with unbiasedness requires Fibonaccian, with a half-golden growth ratio.  We demonstrated the model empirically (functional-form, enumerative and combinatorial properties, 3 predicted rates), and connected sample growth to the statistical environment (biases and predictability) it creates for its populations.

\section*{Declarations}
\textbf{Conflicts of interest/Competing interests}: No conflicts to declare.\\
\textbf{Availability of data and material}: The datasets analysed in the current study are available in the IPUMS repository~\citep{ipums}.\\
\textbf{Funding}: Funding provided by the Sao Paulo Research Foundation (FAPESP).\\
\textbf{Authors' contributions}: AFR is the sole author.\\



\bibliographystyle{plainnat}

\bibliography{bib}

\end{document}